\documentstyle[eqsecnum,aps,epsfig,amssymb]{revtex} 
\def\ii{{\rm i}} 
 
\def\d{\delta} 
\def\D{\Delta} 
\def\l{\lambda} 
\def\o{\omega} 
\def\z{^{(0)}} 
 
\def\th{\vartheta} 
\def\ph{\varphi} 
\def\pa{\partial} 
\def\nn{\nonumber} 
\def\be{\begin{equation}} 
\def\ee{\end{equation}} 
\def\beq{\begin{eqnarray}} 
\def\eeq{\end{eqnarray}} 
\begin{document} 
\title{Non-adiabatic oscillations of compact stars in general relativity} 
\author{ L. Gualtieri$^1$, J.A. Pons $^2$, and G. Miniutti $^3$}  
\address{$^1$ Dipartimento di Fisica ``G.Marconi", 
 Universit\` a di Roma ``La Sapienza"\\ 
and Sezione INFN  ROMA1, piazzale Aldo  Moro 
2, I-00185 Roma, Italy} 
\address{$^2$ Departament de F\'{\i}sica Aplicada, Universitat d'Alacant,\\ 
Apartat de correus 99, 03080 Alacant, Spain} 
\address{$^3$ 
Institute of Astronomy, University of Cambridge, Madingley Road, Cambridge CB3 0HA} 
\maketitle  
\begin{abstract} 
We have developed a formalism to study non-adiabatic, non-radial
oscillations of non-rotating compact stars in the frequency domain, 
including the effects of thermal diffusion in the framework of general
relativistic perturbation theory.  When a general equation of state
depending on temperature is used, the perturbations of the fluid
result in heat flux which is coupled with the spacetime geometry
through the Einstein field equations. Our results show that the
frequency of the first pressure ($p$) and gravity ($g$) oscillation
modes is significantly affected by thermal diffusion, while that of
the fundamental ($f$) mode is basically unaltered due to the global
nature of that oscillation. The damping time of the oscillations is
generally much smaller than in the adiabatic case (more than two
orders of magnitude for the $p-$ and $g-$modes) reflecting the effect
of thermal dissipation. Both the isothermal and adiabatic limits are
recovered in our treatment and we study in more detail the
intermediate regime. Our formalism finds its natural astrophysical
application in the study of the oscillation properties of newly born
neutron stars, neutron stars with a deconfined quark core phase, or
strange stars which are all promising sources of gravitational waves
with frequencies in the band of the first generation and advanced
ground-based interferometric detectors.
\end{abstract} 
\section{Introduction} 
The theory of stellar oscillations has been a fundamental tool in the
study of stellar interiors and stellar properties during decades. For
Newtonian stars the theory is well established and observationally
tested. In some areas such as helioseismology, very high precision
measurements of the normal oscillation frequencies allow for a
detailed understanding of the properties of the solar interior (see
\cite{ChDa} for a review). It is usually assumed that stellar
pulsations are adiabatic because the thermal relaxation timescale in
stellar interiors is orders of magnitude larger than the pulsation
periods. However, in some particular situations, energy transfer in
the external regions of stars is fast enough to affect the pulsation
properties, and some work has been devoted to study, for example,
non-adiabatic oscillations of white dwarfs \cite{LB93} or the coupling
between the different non-linear modes in non--adiabatic situations
\cite{vH94}.  Concerning relativistic stars, the study of pulsation
properties of neutron stars or compact objects such as strange stars
or hybrid stars has become more popular in the last decade (see
e.g. the reviews in Ref. \cite{rev1,rev2,rev3}), mainly due to the
expectations of detecting the gravitational emission from pulsating
nearby compact stars with the current or next generation ground-based
interferometric detectors (LIGO, VIRGO, GEO600, TAMA).  
The theory of non-adiabatic
relativistic stellar pulsations is not as well developed as its
Newtonian cousin, due the higher complexity of the formalisms, but
also due to the fact that most attention has been paid to the study of
old neutron stars, in which the thermal conductivity is too small to
have visible non-adiabatic effects. Nevertheless, there might be
situations in which this is not entirely true. For example, in a newly
born neutron star the thermal structure is determined by neutrino
diffusion \cite{BL86,Pon99}, instead of electron conduction as in old
neutron stars \cite{BHY01}, and the effects of non-adiabaticity are
likely to be relevant in the outer layers. Another interesting
possibility is the existence of deconfined quark matter in neutron
star cores, or in the form of strange stars. There seems to be common
agreement in that, if strange matter exists, it has to be in a color
superconducting phase. It has been recently pointed out \cite{SE02}
that the thermal conductivity of such exotic matter is many orders of
magnitude larger than in normal neutron star matter, and for
sufficiently high temperatures the timescale for thermal relaxation
can be as short as $10^{-4}$ s., comparable with typical oscillation
periods of compact objects.  If this or other exotic scenarios
(hyperons, kaon condensates) happen to be true, our common belief that
compact star pulsations are adiabatic must be modified, and some care
must be taken to understand how non-adiabatic effects change the
oscillation properties and therefore the predicted gravitational wave
signals of pulsating compact stars.
\par 
With this motivation, we have derived a formalism that includes the effects 
of heat transfer and chemical diffusion (important in proto-neutron stars, where 
lepton diffusion is the driving force of thermodynamical changes)  
in a relativistic analysis of stellar 
perturbations. Some work in this line has been done in the past for radial oscillations 
\cite{MMIH}. We have considered the case of non-radial oscillations, since this 
is the case of interest for gravitational wave emission, by extending in a 
simple way the formalism of Lindblom and Detweiler \cite{LD,DL}, and complementing 
the system of equations in the frequency domain with the additional equations  
for thermal or chemical diffusion. 
In the present study we focused on the effects of heat transfer and chemical 
diffusion, neglecting the rotation of the star.
\par 
The paper is structured as follows.  In Sec. II  we derive the equations  
of non-adiabatic stellar perturbations in general relativity. 
In Sec. III we describe the equation of state we have used and we define
the thermodynamical quantities we need in our derivation.
The additional equations (transport of  
energy) that close the system are discussed in section IV.
Section V is devoted to the numerical implementation of the complete set of equations.
In Sec. VI  we discuss in detail the results of the numerical integrations and in
Sec. VII we draw the main conclusions and comment on possible future extensions of 
this work. 
\section{Derivation of the equations} 
The stress--energy tensor of a non--perfect fluid\footnote{
Throughout the paper, we use units in which the Newton gravitational constant, 
the speed of light, and the Boltzmann constant are equal to one, $G=c=k_B=1$.}, 
including heat flux but without viscosity, has the general form \cite{MTW} 
\be 
T_{\alpha\beta} = (\rho+p)u_{\alpha}u_{\beta}+pg_{\alpha\beta} 
+u_{\alpha}q_{\beta}+u_{\alpha}q_{\beta} 
\ee 
where $\rho$ is the energy density, $p$ is the pressure,  $u^{\alpha}$ is the matter  
four--velocity, and $q^{\alpha}$ is the heat flux which satisfies $u_{\alpha}q^{\alpha}=0\,.$ 
In addition, we will also consider the conservation equation for 
the baryon number density $n$ (equation of continuity) 
\be 
\label{cont} 
(nu^{\alpha})_{;\alpha}=0\,. 
\ee 
\subsection{Background configuration}  
Hereafter, we will neglect the heat flux in the background
configuration. This is justified if the background is assumed to be in
thermal equilibrium.  Even in the case that thermal or chemical
gradients are present, the assumption of stationary background is 
valid if the timescale for global thermodynamical changes is much
larger than the timescale of variation of the perturbations. For
example, in newly born neutron stars, structural changes happen on
timescales of the order of 0.1-1 second, while we are interested in
oscillations of periods of the order of milliseconds.  Therefore,
labeling background quantities by the superscript $\z$ , in the
following we assume that \be q_{\mu}\z=0 \ee such that, our background
is a perfect fluid, spherically symmetric star, and is described by
the well known TOV equations: \beq g\z_{\mu\nu}&=&{\rm
diag}\left(-e^{\nu(r)},e^{\l(r)},r^2\gamma_{ab}\right)\\
u^{(0)\,\mu}&=&\left(e^{-\nu/2},0,0,0\right)\label{defu0}\\
T\z_{\mu\nu}&=&(\rho\z+p\z)u\z_{\mu}u\z_{\nu}+p\z g\z_{\mu\nu}\nn\\
\lambda'&=&-\frac{2}{r^2}e^{\l}(M-4\pi\rho\z r^3)\\
\nu'&=&\frac{2}{r^2}e^{\l}(M+4\pi p\z r^3)\\
p^{(0)\prime}&=&-\frac{1}{2}(\rho\z+p\z)\nu_{,r} \eeq where we denote
with a prime $\pa/\pa r$, and we have defined \be
\gamma_{ab}\equiv{\rm diag}\left(1,\sin^2\th\right) \ee (here and in
the following, greek indexes $\mu=0,\dots,3$ run on spacetime, latin
indexes $i=1,\dots,3$ run on the spatial subspace, and latin indexes
$a=\theta,\phi$ run on the sphere).
\subsection{Perturbations} 
The equations of stellar perturbations in the perfect fluid case 
have been derived by many authors in different formalisms \cite{altri,altri2}.  
In this paper, we follow the notation of Lindblom \& Detweiler \cite{LD,DL}, 
(LD hereafter) and we will try to take the parallelism between their  
and our equations as far as possible. 
The perturbed stress--energy tensor has the form  
\beq 
\d T_{\mu\nu}&=&(\d\rho+\d p)u_{\mu}\z u_{\nu}\z+ 
(\rho\z+p\z)(\d u_{\mu}u\z_{\nu}+u\z_{\mu}\d u_{\nu})\nn\\ 
&&+\d p g\z_{\mu\nu}+p\z\d g_{\mu\nu} 
+(\d q_{\mu}u\z_{\nu}+u\z_{\mu}\d q_{\nu})\,.\label{pertT} 
\eeq 
Following the conventions in \cite{LD,DL}, we expand in spherical harmonics the  
metric perturbations with polar symmetry (we will not discuss perturbations  
with axial symmetry, which are not related with fluid oscillations in  
non--rotating stars) as 
\be 
\d g_{\mu\nu}=-r^le^{\ii\o t}\left(\begin{array}{c|c} 
\begin{array}{cc} H_{0\,lm}e^{\nu} Y^{lm} & \ii\o rH_{1\,lm}Y^{lm}\\ 
\ii\o rH_{1\,lm}Y^{lm} & H_{0\,lm}e^{\lambda}Y^{lm} \\ \end{array}  
& 0 \\ 
\hline 
0 & r^2\gamma_{ab}K_{lm}Y^{lm} 
\end{array}\right)\,.\label{metricexpansion} 
\ee 
The Lagrangian displacements are expanded as  
\beq 
\xi^r&=&\frac{e^{-\l/2}}{r}W_{lm}Y^{lm}r^le^{\ii\o t}\nn\\ 
\xi^a&=&-\frac{1}{r^2}V_{lm}\gamma^{ab}Y^{lm}_{,b}r^le^{\ii\o t}\,,
\eeq 
and they are related to the four--velocity perturbation by  
\be 
\d u^i=u^{(0)0}\xi^i_{,t}=\ii\o e^{-\nu/2}\xi^i\,,~~~ (i=1,2,3) 
\ee  
so that its expansion in harmonics is 
\be 
\d u^{\mu}=\left(-\frac{1}{2}e^{-\nu/2}H_{0\,lm}Y^{lm},\, 
\ii\omega\frac{e^{-(\l+\nu)/2}}{r}W_{lm}Y^{lm},\,-\ii\omega\frac{e^{-\nu/2}}{r^2}V_{lm} 
\gamma^{ab}Y_{,b}^{lm}\right)r^le^{\ii\omega t} \,.
\ee 
We also expand the perturbed heat flux in harmonics as follows 
\be 
\delta q_{\alpha}=\kappa e^{-\nu/2} 
\left(0,Q_{1\,lm}(r)r^{l-1}Y^{lm}(\th,\ph) 
,Q_{2\,lm}(r)r^lY^{lm}_{,a}(\th,\ph)\right)e^{\ii\omega t}\,, 
\ee 
($\kappa$ thermal conductivity)  
and the Lagrangian perturbations of the pressure and the energy density as 
\beq 
\Delta p&\equiv&\delta p+p^{(0)}_{,\mu}\xi^{\mu}=\Delta p_{lm}Y^{lm}r^le^{\ii\omega t}\nn\\ 
\Delta \rho&\equiv&\delta\rho+\rho^{(0)}_{,\mu}\xi^{\mu}=\Delta \rho_{lm}Y^{lm}r^l 
e^{\ii\omega t} 
\eeq 
(the same holds for the Lagrangian perturbations of the other thermodynamical quantities). 
\par  
Finally, following and generalizing LD, we will use a set of variables 
($X_{lm},E_{lm}, \Sigma_{lm}$) related to the Lagrangian 
perturbations of the pressure, energy density, and entropy, defined as follows: 
\beq 
X_{lm}&=&-e^{\nu/2}r^{-l}\Delta p_{lm}\nn\\ 
E_{lm}&=&-e^{\nu/2}r^{-l}\Delta \rho_{lm}\nn\\ 
\Sigma_{lm}&=&-e^{\nu/2}r^{-l}\Delta s_{lm}\,.\label{XESY} 
\eeq 
\par 
At this point, we already can see that the presence of the new 
dissipative terms affects the resulting equations in two main 
ways. First, we have a number of additional terms in the perturbed 
energy--stress tensor, related all of them to the heat flux. Second, 
the thermodynamical relation between the pressure and the energy 
density is now more complex because the equation of state (EOS) is not only a 
function of the energy density, but also of the entropy, say, 
$p=p(\rho,s)$. For clarity, we will discuss the above points 
separately. 
\subsubsection{Taking into account the heat flux in the stress--energy tensor.}  
The new terms appearing in the perturbed equations can be greatly simplified 
by working in a suitable reference frame. Introducing the four--velocity of an observer 
in the frame in which the {\it energy} (as opposed to the {\it matter}) is at rest 
\be 
\d\hat u_{\mu}\equiv\d u_{\mu}+\frac{\d q_{\mu}}{\rho\z+p\z}\,,\label{defhatu} 
\ee 
then, the stress--energy tensor (\ref{pertT}) has formally the expression of the  
perfect fluid case: 
\be 
\d T_{\mu\nu}=(\d\rho+\d p)u\z_{\mu}u\z_{\nu}+\d pg\z_{\mu\nu}+p\z\d g_{\mu\nu}+ 
(\d\hat u_{\mu}u\z_{\nu}+\d\hat u_{\nu}u\z_{\mu})\,.\label{standardtmunu} 
\ee 
Accordingly, we can redefine the original variables 
of Lindblom \& Detweiler in the following way: 
\beq 
\hat W_{lm}&\equiv&W_{lm} 
+e^{-\l/2}\frac{\kappa}{\ii\omega}\frac{Q_{1\,lm}}{\rho\z+p\z}\label{Wh}\\ 
\hat V_{lm}&\equiv&V_{lm} 
-\frac{\kappa}{\ii\omega}\frac{Q_{2\,lm}}{\rho\z+p\z}\label{Vh}\\ 
\hat X_{lm}&\equiv&X_{lm} 
-e^{\nu/2-\l}\frac{\kappa p^{(0)\prime}}{\ii\omega r}\frac{Q_{1\,lm}}{\rho\z+p\z}\label{Xh}\\ 
\hat E_{lm}&\equiv&E_{lm} 
-e^{\nu/2-\l}\frac{\kappa\rho^{(0)\prime}}{\ii\omega r}\frac{Q_{1\,lm}}{\rho\z+p\z}\,, 
\label{Eh} 
\eeq 
and now the harmonic expansions of $\d u_{\mu}$ and of the thermodynamical quantities  
are also formally identical to those in the perfect fluid case: 
\beq 
\d\hat u^{\mu}&=&\left(-\frac{1}{2}e^{-\nu/2}H_{0\,lm}Y^{lm},\, 
\ii\omega\frac{e^{-(\l+\nu)/2}}{r}\hat W_{lm}Y^{lm},\, 
-\ii\omega\frac{e^{-\nu/2}}{r^2}\hat V_{lm}\gamma^{ab}Y_{,b}^{lm}\right)r^le^{\ii\omega t} 
\nn\\ 
&&\\ 
\hat X_{lm}&=& 
-e^{\nu/2}\left(r^{-l}\delta p_{lm} 
+\frac{e^{-\l/2}}{r}\hat W_{lm}p^{(0)\prime}\right)\\ 
\hat E_{lm}&=& 
-e^{\nu/2}\left(r^{-l}\delta\rho_{lm} 
+\frac{e^{-\l/2}}{r}\hat W_{lm}\rho^{(0)\prime}\right) 
\,. 
\eeq 
Thus, we can follow the formal derivation of the perturbation equations in 
the adiabatic case, but using variables in the energy frame,  the {\it hatted} 
variables; in this way we take easily into account the extra terms in the 
stress--energy tensor. The final system of equations (see section IV) will be 
analogous to the original system in the formalism of LD, if the perturbation 
of the energy $\hat E$ is left explicitly in the equations. 
\par  
\subsubsection{Relation between $\Delta\rho$ and $\Delta p$} 
\par 
In the adiabatic case, the system of equations is closed by using the EOS to 
express the energy perturbation in terms of the pressure perturbation, 
{\it i.e.} $ \D\rho=c_s^{-2}\D p $ 
where $c_s^{2}$ is the adiabatic speed of sound 
\be 
c_s^{2}\equiv\left(\frac{\partial p}{\partial \rho}\right)_{s}\,. 
\ee 
This relation is used in the derivation of the LD equations and by many other 
authors in order to eliminate the  $\D\rho$ from the equations, so that in the  
final system only the pressure perturbation appears. 
 
For general EOSs (of the form, i.e., $\rho=\rho(p,s)$),  
the Lagrangian perturbation of the energy density can 
be expressed in terms of the  Lagrangian perturbations of the other variables 
by using the thermodynamical relation: 
\be 
d\rho=c_s^{-2}dp+nT\alpha_1ds\,,\label{relalpha} 
\ee 
where $T$ is the matter temperature, $n$ is the particle density 
(baryon density for our purposes in neutron stars) and 
\beq 
\alpha_1&\equiv&\frac{1}{nT}\left(\frac{\pa\rho}{\pa s}\right)_{p}\nn. 
\eeq 
Therefore, the Lagrangian perturbation of the energy density is given by 
\be 
\D\rho=c_s^{-2}\D p+n\z T\z\alpha_1\D s\,.
\label{Deltarho2} 
\ee 
In the following we will omit the $\z$ super-indexes that label background  
quantities ($n, T, \rho, p$) for clarity.  More details about the equation of state and  
thermodynamics are given in the next section. 
\par  
The perfect--fluid limit is recovered when the element of fluid does not  
exchange heat with its surroundings, i.e., $\Delta s=0$. 
In the general case, $\Delta s$ does not vanish, so we have to take 
into account all terms in equation (\ref{Deltarho2}), which couples 
the entropy perturbation to the other perturbations.  Therefore, instead of 
simply substituting $E_{lm}=c_s^{-2} X_{lm}$, one must now use 
\be 
\hat E_{lm}=c_s^{-2}\hat X_{lm}+n T \alpha_1\hat\Sigma_{lm} \,.\label{exprhE} 
\ee 
But this introduces a new variable, the perturbation of the entropy,
so that we need additional information.
\par 
\subsubsection{Using the energy and baryon conservation equations} 
The new variable which is now included in our system of equations can be 
rearranged by using the first law of thermodynamics and the continuity equation, 
which was not necessary in the perfect fluid case. 
The time component of the divergence of the stress--energy tensor,  
i.e. the energy density conservation equation, up to first order in 
the perturbations, reads 
\be 
u_{\alpha}T^{\alpha\beta}_{;\beta}= 
-\ii\omega e^{-\nu/2}\Delta\rho-(\rho+p)u^{\beta}_{;\beta} 
-\left(\delta q^{\beta}_{~;\beta}+\delta q^{\alpha}u^{\beta} 
u_{\alpha;\beta}\right)=0\,. 
\ee 
In order to simplify this expression we can use the equation of continuity 
(\ref{cont}) 
\be 
(nu^{\mu})_{;\mu}=u^{\mu}n_{, \mu}+nu^{\mu}_{;\mu}=e^{-\nu/2}\ii\omega\Delta n 
+nu^{\mu}_{;\mu}=0 
\ee 
and the first principle of thermodynamics in the form 
\be 
\Delta \rho=\frac{\rho+p}{n} \Delta n+nT \Delta s\,, 
\ee 
to obtain 
\be 
{\ii\omega} n T\Delta s=-{e^{\nu/2}}
\left(\delta q^{\beta}_{~;\beta}+\delta q^{\alpha}u^{\beta}u_{\alpha;\beta}\right) 
\label{peqSY1} 
\ee 
that\footnote{ 
Notice that when $\delta q^{\alpha}=0$, equation (\ref{peqSY1}) gives 
$\Delta s=0$: the adiabaticity of the perturbations directly follows 
from the form of the perfect fluid stress--energy tensor.}, 
in terms of $\hat\Sigma_{lm}$, becomes 
\beq 
&&{\ii\omega} n T \hat\Sigma_{lm}=\kappa\frac{e^{\nu/2-\l}}{r} 
\left[Q_{1\,lm}'+\left(\frac{l+1}{r}+\frac{\nu'-\lambda'}{2}+\frac{\kappa'}{\kappa} 
-\frac{n T s^{\prime}}{\rho+p}\right) 
Q_{1\,lm}\right.\nn\\ 
&&\left.-e^{\l}\frac{l(l+1)}{r}Q_{2\,lm}\right]\,.\label{eqSY1} 
\eeq 
\par  
Equation (\ref{eqSY1}) gives the entropy perturbation in terms of the 
perturbed energy flux $Q_1$.  In order to close the system, we need to 
supplement our system with the equations that describe how the heat 
flux depends on the local gradients of the perturbations: the 
transport equations, which are given by the relativistic theory of 
dissipative thermodynamics.  
\section{A simple two--parameter Equation of State.} 
In this section we describe the EOS that 
we have used and we define all the thermodynamical quantities  
that appear in the final form of the equations.  We begin by  
considering a general equation of state depending on the baryon 
density $n$ and the entropy per baryon $s$: 
\be 
p=p(n,s)~; ~ 
\rho=\rho(n,s)~.
\ee 
Let us define the specific internal energy $e$  
(energy per particle, excluding the rest mass) and the 
speed of sound $c_s$: 
\beq 
e &=& \frac{\rho}{n} - m_B \\ 
c_s^2 &=& \left(\frac{\pa  p}{\pa \rho}\right)_s 
\eeq 
($m_B$ baryon mass).  
Next, let us assume that the EOS can be written as follows 
\be 
p(n,e) = (\gamma-1)~n~e \label{perfect0} 
\ee 
with $\gamma$ constant. Using thermodynamical identities,  
one can show that  
$\left(\frac{\partial p}{\partial n}\right)_s=\frac{\gamma p}{n}$,  
so that (\ref{perfect0}) is equivalent to 
\be 
p(n,s)=K(s)(m_B n)^\gamma\label{perfect} 
\ee 
where $K(s)$ is an arbitrary function of integration that only depends 
on the specific entropy. If one chooses $K(s)=K_0 \exp{(\gamma-1)s}$, 
this leads to the perfect gas law $p=n T$ (in units with the  
Boltzmann constant $k_B=1$). 
\par  
The form (\ref{perfect}) allows to include finite temperature effects with 
a simple equation of state, just giving an explicit dependence of $K(s)$ with 
the entropy.  From the first law of thermodynamics 
\be 
d\rho = \frac{\rho+p}{n} dn + n T ds \, ,\label{Ilaw} 
\ee 
the temperature can be found according to 
\be 
T=\frac{1}{n}\left(\frac{\partial\rho}{\partial s}\right)_n 
=\frac{1}{n(\gamma-1)}\left(\frac{\partial p}{\partial s}\right)_n 
=\frac{p}{(\gamma-1)n}\frac{d \ln K}{ds}\,.\label{defT} 
\ee 
The rest of thermodynamical quantities we 
need, in terms of $K$, are the following: 
\beq 
\rho&=&\left(\frac{p}{K}\right)^{\frac{1}{\gamma}}+\frac{p}{\gamma-1}\\ 
c_s^2&=&\left[\frac{1}{\gamma K} 
\left(\frac{p}{K}\right)^{\frac{1}{\gamma}-1}+\frac{1}{\gamma-1}\right]^{-1}\\ 
\alpha_1&=&\frac{1}{n T}\left(\frac{\pa\rho}{\pa s}\right)_p 
=1-\frac{\gamma-1}{c_s^2}\\ 
C_p&=&T\left(\frac{\pa s}{\pa T}\right)_p= 
\left[  \frac{d}{ds}\ln\left(\frac{d}{ds} K(s)^{1/\gamma} \right) \right]^{-1} 
\eeq 
where $C_p$ is the specific heat at constant pressure. 
\par  
In order to study the effects of heat transport in the non--adiabatic case, 
we want to give a simple form of EOS which depends non trivially 
on the entropy profile and such that, consistently with the assumption 
that the background flux is negligible,  
it is isothermal in the relativistic sense ($T e^{\nu/2}$=const.)\footnote{ 
We stress that, as we said in the previous section, 
the equations we derive do apply also for background profiles not perfectly 
isothermal, as long as the background evolution timescale is longer than the dynamical
timescale.}.  
This can be done as follows. By using the first law of thermodynamics (\ref{Ilaw})  
it is easy to prove that 
\be 
\frac{dT}{T} = \frac{ds}{C_p} + (1-\alpha_1) \frac{dp}{\rho+p}\,.\label{dToverT} 
\ee 
If we look for relativistic isothermal profiles, this implies that 
\be 
\frac{dT}{T} = \frac{d\nu}{2} = \frac{dp}{\rho+p} 
\ee 
where the last equality comes from the TOV equation (hydrostatic equilibrium). 
Therefore, we have 
\be 
\frac{ds}{C_p} = \alpha_1 \frac{dp}{\rho+p} 
= - \frac{\gamma-1}{\gamma} \left( 1 - \frac{c_s^2}{\gamma-1} \right) \frac{dp}{p} 
\ee 
and, substituting the explicit expression of the speed of sound, 
\be 
\frac{ds}{C_p} = - \frac{\gamma-1}{\gamma} 
\frac{1}{1+\frac{\gamma p}{\gamma-1} \left(\frac{K}{p}\right)^{1/\gamma}} 
\frac{dp}{p}. 
\ee 
Defining $x= \ln{p^{1-1/\gamma}}$, we finally obtain 
\be 
\frac{dx}{ds} = -\left[ 1 + \frac{\gamma}{\gamma-1}e^x f(s) \right] 
\frac{d}{ds}\ln\left[\frac{d}{ds} f(s) \right] 
\ee 
where $f(s)= K(s)^{1/\gamma}$. In principle this differential equation is not separable, 
but by numerical integration we can find the entropy profile $s(p)$ that leads 
to isothermality in the relativistic sense, for any given function $f(s)$. 
In particular, we have chosen   
\be 
f(s) = C + \beta e^{s/a} 
\ee 
that gives a constant specific heat $C_p=a$. Here, $C$ and $\beta$ are arbitrary 
constants fixed to reproduce similar masses, radii, and temperatures  
as for newly born neutron stars. In particular, the temperature of the
model that we analyze in the results section varies between (5-8) $\times 10^{11} K$,
and $C_p=0.5$.
\section{Dissipative relativistic fluids and heat transfer.} 
The theory of standard irreversible thermodynamics was first extended from 
Newtonian to relativistic by Eckart in 1940. 
This theory shares with its Newtonian counterpart the problem that 
perturbations propagate at infinite speeds, which results in unstable 
equilibrium states.  We must notice, however, that the non--causal Newtonian theory has 
been used for years in the study of non--relativistic stars with remarkable 
success, and with observational confirmation. Therefore, there is no reason 
to believe that the standard theory cannot be used as a first order approach 
to the problem and can give reasonable estimates of the main effects, even 
for relativistic stars. 
In general relativity, causality can be restored 
within the framework of an extended theory developed by Israel and Stewart \cite{IS}. 
This extended theory is also known as causal thermodynamics 
or second order thermodynamics because of the appearance of second order 
terms of the dissipative variables in the entropy.  The problems associated  
to non--causal heat transfer become more relevant when the mean free path 
becomes larger than the spatial scale of the problem (in our case, the 
radius of the star), that is, for large thermal conductivities or 
diffusivity. The extended theory automatically incorporates transient 
phenomena on the timescale of the mean free path that cure the inconsistency. 
In this paper, we are interested in understanding qualitatively the 
effects of dissipative terms on the oscillation properties of stars, rather 
than developing a fully consistent theory, which is a more ambitious and complex task. 
Nevertheless, we start with the extended theory of dissipative fluids, 
and later on we discuss which second order terms are neglected and their relative 
importance. 
\par  
We begin with the equation that describes the variation of the entropy  
\beq 
nT u^{\alpha} s_{,\alpha}=-q^{\alpha}_{~;\alpha}-q^{\alpha}u^{\beta}u_{\alpha;\beta}\label{ek1}~,
\eeq 
and the definition of the heat flux, including the relaxation term that restores causality;  
this is of covariant Maxwell--Cattaneo form (a truncated version of Israel--Stewart equations): 
\beq 
 \tau_1 h_{\alpha}^{\beta} u^{\gamma} q_{\beta;\gamma} + q_{\alpha} 
=-\kappa(h_{\alpha}^{\beta} T_{,\beta}+Tu^{\beta}u_{\alpha;\beta})\label{ek2} 
\eeq 
where $h_{\alpha}^{\beta}$ is the projector onto the spatial 
subspace, $h_{\alpha}^{\beta}=\delta_{\alpha}^{\beta}+u_{\alpha}u^{\beta}$, 
$\kappa$ is the thermal conductivity, and $\tau_1$ is the relaxation timescale in which 
the system restores thermal equilibrium. 
Let us consider equations (\ref{ek1}), (\ref{ek2}) on a perturbed 
static spherical background. 
The perturbation of equation (\ref{ek1}) gives the equation previously 
derived (\ref{eqSY1}). The perturbed version of equation (\ref{ek2}) must  
satisfy $q_{\alpha}u^{\alpha}=0$ and, since we are imposing a zero 
background flux ($q\z_{j}=0,~j=1,2,3$), we have $\d q_0=0$ and 
\beq 
(1 + \ii\omega e^{-\nu/2} \tau_1) \d q_{j}&=&-\kappa\left(\delta T_{,j}+\delta T 
u^{(0)\mu}u\z_{j;\mu}+T\z\delta(u^{\mu}u_{j;\mu})\right) 
\,.\label{pert} 
\eeq  
Expanding in spherical harmonics and after some manipulations we find 
\beq 
(1 + \ii\omega e^{-\nu/2} \tau_1) \delta q_{j} 
&=&-e^{-\nu/2}\kappa\left[\delta \left(Te^{\nu/2} 
\right)\right]_{,j}-\kappa T \omega^2e^{-\nu}v_{j}\nn\\ 
&=&-e^{-\nu/2}\kappa\left[\Delta \left(Te^{\nu/2}\right) 
\right]_{,j}-\kappa T \omega^2e^{-\nu}v_{j} 
\label{ekpert} 
\eeq 
where 
\be 
v_{j}\equiv\left(\left[rH_{1\,lm}-\frac{e^{\lambda/2}}{r}W_{lm} 
\right]Y^{lm},V_{lm}Y^{lm}_{,a}\right)r^le^{\ii\omega t}\,.
\label{defvj} 
\ee 
Applying the thermodynamical relation (\ref{dToverT}), which implies  
\be 
\frac{\Delta T}{T} = \frac{\Delta s}{C_p} + (1-\alpha_1) \frac{\Delta p}{\rho+p}\,, 
\ee 
on Eq. (\ref{ekpert}), changing to the energy frame (perturbations in terms  
of hatted variables), and applying the Einstein equation  
\be 
\hat X_{lm}-\omega^2(\rho+p)e^{-\nu/2}\hat V_{lm}-\frac{\nu'}{2r}e^{(\nu-\lambda)/2} 
(\rho+p)\hat W_{lm}-\frac{1}{2}(\rho+p)e^{\nu/2}H_{0\,lm}=0\,,\label{einsteineqnew} 
\ee 
we obtain a first algebraic relation between 
the entropy perturbation, $Q_2$, and $\hat X$, 
\be 
(1 + \ii\omega e^{-\nu/2} \tau_1) Q_{2\,lm}=T\left[\frac{\hat\Sigma_{lm}}{C_p} 
-\frac{\alpha_1}{\rho+p}\hat X_{lm} 
+\frac{\ii\omega\kappa e^{-\nu/2}}{\rho+p}Q_{2\,lm}\right]\,.\label{sigmainv0} 
\ee 
The exact evaluation of the relaxation time $\tau_1$ involves complicated collision 
integrals and depends on the microscopical details of the interaction between 
particles. Notice though, that it is usually estimated as  
$\approx \kappa T/ \rho$ (see i.e. \cite{IS}).  
Therefore we do know that this second order 
correction is of the same order as the last term on the right hand side of 
Eq. (\ref{sigmainv0}). Indeed, if one defines the relaxation timescale as 
$\tau_1 = \kappa T / (\rho+p)$, these two terms cancel each other. 
In the following, we will neglect all second order terms, and Equation \ref{sigmainv0} 
becomes 
\be 
Q_{2\,lm}=T\ \left[\frac{\hat\Sigma_{lm}}{C_p} 
-\frac{\alpha_1}{\rho+p}\hat X_{lm} 
\right]\,.\label{sigmainv} 
\ee 
We have checked that the numerical results presented in section VI are not 
altered by including the second order corrections, being the relative difference 
between the two solutions less than $1\%$. 
\par  
Let us come back to Eq. (\ref{ekpert}). After neglecting the second order term, 
the quantity $e^{\nu/2}\d q_{j}/\kappa+T \o^2e^{-\nu/2}v_{j}$ is  
a gradient, thus the harmonic components of the perturbed heat flux satisfy 
\beq 
Q_{2\,lm}'&=&-\frac{l}{r}Q_{2\,lm}+\frac{1}{r}Q_{2\,lm} 
\nn\\ 
&&-T\omega^2e^{-\nu/2}\left[\left(\partial_r-\nu'+\frac{l}{r}\right)\hat V_{lm} 
-rH_{1\,lm}+\frac{e^{\lambda/2}}{r}\hat W_{lm}\right]\,.\label{eqQ20} 
\eeq 
Next, we can make use of the following equation, that is derived directly 
from the LD equations \cite{LD,DL} 
\beq 
(\omega^2e^{-\nu}\hat V_{lm})' = 
 \omega^2e^{-\nu}\left(rH_{1\,lm}-\frac{e^{\lambda/2}}{r}\hat W_{lm}-\frac{l}{r} 
\hat V_{lm}\right) - \frac{e^{-\nu/2}}{(\rho+p)^2}(\hat X_{lm}\rho^{\prime} 
-\hat E_{lm}p^{\prime}) 
\eeq 
to simplify the right hand side of Eq. (\ref{eqQ20}) 
\beq 
Q_{2\,lm}'&=&-\frac{l}{r}Q_{2\,lm}+\frac{1}{r}Q_{1\,lm} 
+\frac{T}{(\rho+p)^2}(\hat X_{lm}\rho'-\hat E_{lm}p') 
~,\label{Q2parz} 
\eeq 
and, finally, introducing the thermodynamical derivatives defined
in the previous section ($\alpha_1, C_p$)  
and keeping only first order terms we obtain  
\be 
Q_{2\,lm}'=-\frac{l}{r}Q_{2\,lm}+\frac{1}{r}Q_{1\,lm} 
+nT\alpha_1 \frac{\nu'}{2}\frac{C_p}{(\rho+p)}Q_{2\,lm}\,. 
\ee 
\section{Numerical implementation} 
\subsection{The complete set of equations} 
We can now write the full set of equations. For simplicity, in the following 
we will omit the sub-indexes ${lm}$, keeping in mind that one has a complete 
set of equations for each multipole in the expansion. 
Our variables are  
\be 
H_{0},\,H_{1},\,\hat K,\,\hat W,\,\hat V,\,\hat X,\, 
\hat E,\,\hat\Sigma,\,Q_{1},\,Q_{2}\,. 
\ee 
Four of these variables 
$H_{0},\,\hat V,\,\hat E,\,\hat\Sigma,$ 
are given in terms of the others by the following algebraic relations: 
\beq 
H_{0}&=&\frac{1}{3M+\frac{(l-1)(l+2)}{2}r+4\pi r^3p}\left\{ 
8\pi r^3e^{-\nu/2}{\hat X} 
- r^3 e^{-\l} \left[\frac{l(l+1)}{2}\frac{\nu'}{2 r}-\omega^2 e^{-\nu}\right]H_{1} 
\right.\nn\\ 
&& \left. 
+\left[\frac{(l-1)(l+2)}{2}r 
- \omega^2r^3e^{-\nu}-\frac{r^2 \nu'}{2}e^{-\l} \left(\frac{r \nu'}{2}-1 \right)\right] 
K\right\}\nn\\ 
&&\label{exH0}\\ 
&&\nn\\ 
\omega^2 e^{-\nu} \hat V&=& \frac{\hat X e^{-\nu/2}}{(\rho+p)} 
-\frac{e^{-\l/2}}{r}\frac{\nu^{\prime}}{2}{\hat W}- 
\frac{1}{2} H_{0} \label{exV}\\ 
&&\nn\\ 
\hat E&=&c_s^{-2}\hat X+n T\alpha_1\hat\Sigma \label{exE}\\ 
&&\nn\\ 
\hat\Sigma&=&C_p\left[\frac{Q_2}{T} 
+\frac{\alpha_1}{\rho+p}\hat X  
\right]\,. 
\label{exSigma} 
\eeq  
The remaining variables 
$ H_{1},\,K,\,{\hat W},\,{\hat X},\,Q_{1},\,Q_{2},\,$ 
satisfy the following differential equations: 
\beq 
H_{1}'&=&-\frac{1}{r}\left[l+1+2M\frac{e^{\l}}{r}+4\pi r^2e^{\l}(p-\rho)\right] 
H_{1}\nn\\ 
&&+\frac{1}{r}e^{\l}\left[H_{0}+K-16\pi(\rho+p){\hat V}\right]\label{eqH1}\\ 
&&\nn\\ 
K'&=&\frac{1}{r}H_{0}+\frac{l(l+1)}{2r}H_{1} 
-\left[\frac{l+1}{r}-\frac{1}{2}\nu'\right]K 
-8\pi(\rho+p)\frac{e^{\l/2}}{r}{\hat W}\label{eqK}\\ 
&&\nn\\ 
{\hat W}'&=&-\frac{l+1}{r}{\hat W}+re^{\l/2} 
\left[\frac{e^{-\nu/2}}{\rho+p}{\hat E}- 
\frac{l(l+1)}{r^2}{\hat V}+\frac{1}{2}H_{0}+K\right]\label{eqW}\\ 
&&\nn\\ 
{\hat X}'&=&-\frac{l}{r}{\hat X}+(\rho+p)e^{\nu/2}\left\{\frac{1}{2}\left( 
\frac{1}{r}-\frac{1}{2}\nu'\right)H_{0}\right.\nn\\ 
&&+\frac{1}{2}\left[r\omega^2e^{-\nu}+\frac{l(l+1)}{2r}\right]H_{1} 
+\frac{1}{2}\left(\frac{3}{2}\nu'-\frac{1}{r}\right)K 
-\frac{l(l+1)}{2r^2}\nu'{\hat V}\nn\\ 
&&\left. 
-\frac{1}{r}\left[4\pi(\rho+p)e^{\l/2}+\omega^2e^{\l/2-\nu} 
-\frac{r^2}{2}\left(\frac{e^{-\l/2}}{r^2}\nu'\right)'\right]{\hat W}\right\}\label{eqX}\\ 
&&\nn\\ 
Q_2'&=&-\frac{l}{r}Q_2+\frac{1}{r}Q_1 
-nT\alpha_1p'\frac{C_p}{(\rho+p)^2}Q_2 
\label{eqq2}\\ 
&&\nn\\ 
Q_1'&=&nT\frac{\ii\omega re^{\lambda-\nu/2}}{\kappa}\hat\Sigma 
-\left(\frac{l+1}{r}+\frac{\nu'-\lambda'}{2}+\frac{\kappa'}{\kappa} 
-\frac{nTs'}{\rho+p}\right)Q_1+e^{\lambda}\frac{l(l+1)}{r}Q_2\,.\label{eqq1} 
\eeq 
It must be noticed that the first four equations are formally identical to those 
of Lindblom \& Detweiler \cite{LD,DL}, but in terms of the new 
variables defined by Eq. (\ref{Wh})--(\ref{Eh}).   
\subsection{Boundary conditions} 
We seek for solutions of the perturbation 
equations which are regular at the origin. Assuming that all variables 
near the center of the star have the form $x(r)=x(0)+O(r^2)$, and 
expanding equations (\ref{exH0})--(\ref{eqq1}) we find 
that the following relations must be satisfied at the origin: 
\beq 
H_{0}(0)&=&K(0)\label{H00}\\ 
{\hat V}(0)&=&-\frac{1}{l}{\hat W}(0)\label{V0}\\ 
{\hat X}(0)&=&(\rho_0+p_0)e^{\nu_0/2} 
\left\{\left[\frac{4\pi}{3}(\rho_0+3p_0) 
-\omega^2\frac{e^{-\nu_0}}{l}\right]{\hat W}(0)+\frac{1}{2}K(0)\right\}\nn\\ 
&&\label{X0}\\ 
H_{1}(0)&=&\frac{1}{l(l+1)}\left[2lK(0)+16\pi(\rho_0+p_0) 
{\hat W}(0)\right]\label{H10}\\ 
Q_{2}(0)&=&\frac{1}{l}Q_{1}(0)\,.\label{q20} 
\eeq 
The values of $\hat E_{lm}$ and $\hat\Sigma_{lm}$  
at the origin are given by (\ref{exE}), (\ref{exSigma}) evaluated at $r=0$. 
Therefore, our equations admit three independent solutions, which are
reduced to one by imposing appropriate boundary conditions at the surface of 
the star. The first is the vanishing of the Lagrangian pressure perturbation 
\be 
X(R_s)=0\,;\label{Xs0} 
\ee 
and the second is the vanishing of the radial flux
\be 
Q_1(R_s)=0\,. 
\ee 
Once we know the only independent solution at the surface of the star,  
the Zerilli function and its derivative can be computed in terms of 
the $H_1$ and $K$, as usual (see for example \cite{IP} and \cite{koj} 
for the expression of the Zerilli function in terms of $H_0$, $K$; the 
expression in terms of $H_1$ follows by applying Eq.(\ref{exH0}) ): 
\beq 
Z&=&r^l\frac{2r^2}{(l-1)(l+2)r+6M}(K-e^{\nu}H_{1})\nn\\ 
Z'&=&r^l\left(\frac{2(l-1)(l+2)r^2-6M((l-1)(l+2)r+2M)}{((l-1)(l+2)r+6M)^2}K 
\right.\nn\\ 
&&\left.+e^{2\nu}\frac{(l-1)l(l+1)(l+2)r^2+6M((l-1)(l+2)r+4M)}{((l-1)(l+2)r+6M)^2} 
H_{1}\right)\,.\nn\\ 
\eeq 
Notice that the metric variables $H_0, H_1, K$ had not been 
redefined by the presence of heat flux.  
Finally, we integrate the Zerilli equation outside the star 
using the continued fraction method (see e.g. \cite{Sotani,gdisc}), 
to obtain the amplitude 
of the ingoing and outgoing parts of the gravitational wave. The quasi-normal 
modes will correspond to the solutions for which the wave is purely outgoing. 
\begin{figure}[htb] 
\epsfxsize=14cm \epsfbox{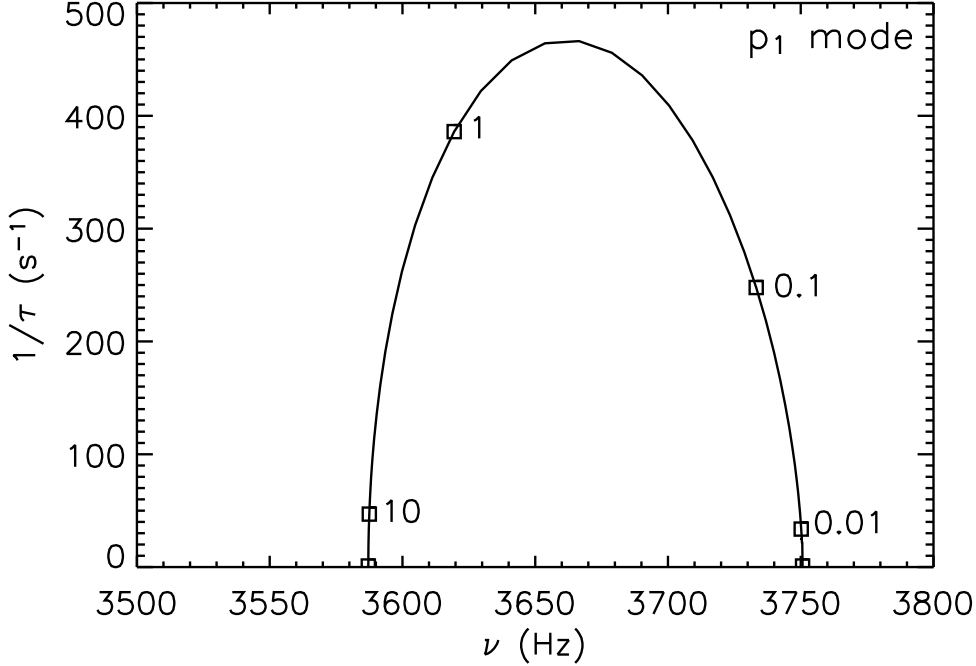} 
\caption{Representation, in the complex plane, of the variation of the
frequency and inverse damping time (imaginary part of the frequency) as the coefficient
$\kappa/n$ varies from the adiabatic limit ($\ll R$) to the isothermal limit ($\gtrsim R$).
The results correspond to the first $p-$mode. As expected, the thermal damping is maximum
when $\kappa/n$ is of the order of unity.}
\label{figp} 
\end{figure} 
\section{Results} 
In the previous sections we have ignored the effects of chemical 
diffusion. Our equations can then be applied to a physical situation 
in which heat conduction is relevant, while chemical diffusion is not. 
This could be the case of a young, hot, neutron star 
about 30 seconds after birth, once the proto-neutron star has lost its lepton content, and 
the temperature gradient is not far from isothermal. 
It is also the situation that one expects to find in a cold neutron star 
that undergoes a phase transition to deconfined quark matter in a color 
superconductor state.  In this latter scenario, pairing between quarks 
inhibits the presence of leptons \cite{AR02}. 
For reference, in Appendix \ref{nm} we derive the equations including 
the terms related to chemical diffusion, leaving for future work their 
application in a realistic scenario. In this paper, we will focus only on 
the effects of finite thermal conductivity, from which we can understand 
most of the qualitative differences between the adiabatic and non--adiabatic cases. 
Hereafter we consider a profile for the thermal conductivity such that
$\kappa/n$ is constant throughout the star. The ratio $\kappa/n$ represents, roughly,
a sort of mean free path, to be compared with the characteristic scale of the system,
i.e., the radius of the star. In the diffusion limit $\kappa/n \ll R$, perturbations
can be considered basically adiabatic, while when $\kappa/n$ becomes of the order or
larger than the radius of the star, thermal equilibrium is achieved in a timescale
shorter than the dynamical characteristic timescale.
\par 
We begin by studying the effects of finite thermal conductivity on the
$p-$modes.  In Fig. \ref{figp}, we show, in the complex plane, the
variation of the complex frequency as the ratio $\kappa/n$ varies from
the adiabatic limit to the isothermal limit.  The oscillation
frequency (real part) is represented in the horizontal axis while the
vertical axis corresponds to the inverse damping time (imaginary
part).  From the results, we can see that the real part of the
frequency of the first acoustic mode is shifted to lower values in
about 200 Hz, as the mean free path becomes larger, an the isothermal
limit is approached. The damping time in both, the adiabatic and
isothermal limit is very similar (not visible in the scale of the
figure), being in both cases of about 1 s. However, in the
semi-transparent regime, thermal dissipation is so effective that the
damping time is reduced by about 3 orders of magnitude (to 2-3 ms).
These results can be easily understood by looking at a
simple toy model that describes how acoustic modes are affected by
thermal diffusion. We describe this simple Newtonian problem in
Appendix \ref{toy}. It makes more manifest the relevant physics
without the complications of the full set of equations in General
Relativity. By analogy with this toy model, one can understand that
the oscillation frequency in the isothermal limit is smaller than that
of the adiabatic case, depending on the ratio between the adiabatic
speed of sound $c_s$ and the isothermal speed of sound $c_T$, defined
as \be c_T^2\equiv\left(\frac{\partial p}{\partial \rho}\right)_T =
\left( \frac{1}{c_s^2} + \frac{n T \alpha_1 C_p (\alpha_1-1)}{\rho+p}
\right)^{-1}\,.  \ee For the EOS we used, this difference is about a
5\% in average, which is consistent with the shift in the $p-$mode
frequency.
\par 
Several comments about the adiabatic and the isothermal limits are in order. 
In both cases, the heat flux tends to zero as we approach the limits, 
but for different reasons, that can be understood by looking at 
Eq. (\ref{sigmainv}). In the diffusion limit (adiabatic perturbations) 
 $\kappa \rightarrow 0$, but in the isothermal limit, the flux vanishes 
because the term within brackets (i.e. the perturbation of the temperature, 
including relativistic corrections) vanishes.  
Therefore, in both limits Eqs. (\ref{eqq2})--(\ref{eqq1}) are decoupled from 
the rest of the system. 
The difference between both limits appears only in Eq. (\ref{eqW}). 
Explicitly, in the adiabatic limit Eq. (\ref{eqW}) reads: 
\be 
{\hat W}'=-\frac{l+1}{r}{\hat W}+re^{\l/2} 
\left[\frac{e^{-\nu/2}}{\rho+p}\frac{\hat X}{c_s^2} - 
\frac{l(l+1)}{r^2}{\hat V}+\frac{1}{2}H_{0}+K\right] 
\ee 
while in the isothermal limit it becomes 
\be
{\hat W}'=-\frac{l+1}{r}{\hat W}+re^{\l/2} 
\left[\frac{e^{-\nu/2}}{\rho+p} \frac{\hat X}{c_T^2}
- \frac{n T \alpha_1 C_p}{\rho+p} \frac{\hat{\Delta \nu}}{2}
- \frac{l(l+1)}{r^2}{\hat V}+\frac{1}{2}H_{0}+K\right]
\ee
where 
\be
\hat{\Delta\nu} = H_0 + \frac{e^{-\l/2}\nu'}{r}\hat{W}\,.
\ee
In other words, one just needs to substitute 
the adiabatic speed of sound by the isothermal 
speed of sound (consistently with what happens in 
Newtonian perturbation theory) but now there appears 
an additional term (proportional to the Lagrangian perturbation of
$\nu/2$) because of the fact that the concept of isothermality in general
relativity involves the red-shifted temperature, $T e^{\nu/2}$. 
\par  
The damping time when dissipation is most effective,
i.e. (as we can see from Fig.\ref{figp}) 
when $\kappa/n \approx 1$ km, can be also estimated by
\be 
\tau_{diss}= \frac{s}{ds/dt}= n C_p \frac{R^2}{\kappa},~ 
\label{ediss}
\ee 
and taking typical values at the interior of the star, such as 
$C_p=1$, $R=10$ km, we get that 
for $\kappa/n=1$ km, the dissipative timescale is 
$\tau_{diss}=0.001$ s, in agreement with the results shown in Fig. \ref{figp}.
These results show explicitly and quantitatively that thermal dissipation
affects the damping of the non-radial pulsations, competing with the other
main dissipation mechanism, i.e., GW emission. The damping times we compute
take for the first time into account both mechanisms in a self--consistent
manner, by including the appropriate physics in the equations.
In the semitransparent regime, when diffusion is more effective,
we must expect short-lived, strongly damped, GW signals, more similar to a GW burst
than a proper $\sim$ kHz oscillation lasting for several hundred oscillations.
Notice, however, that if the physical conditions are such that the thermal
relaxation timescale becomes shorter than the oscillation period, thermal damping
does not affect much the damping time, being the only remarkable difference
a shift in the oscillation frequency.
\begin{figure}[htb] 
\epsfxsize=14cm \epsfbox{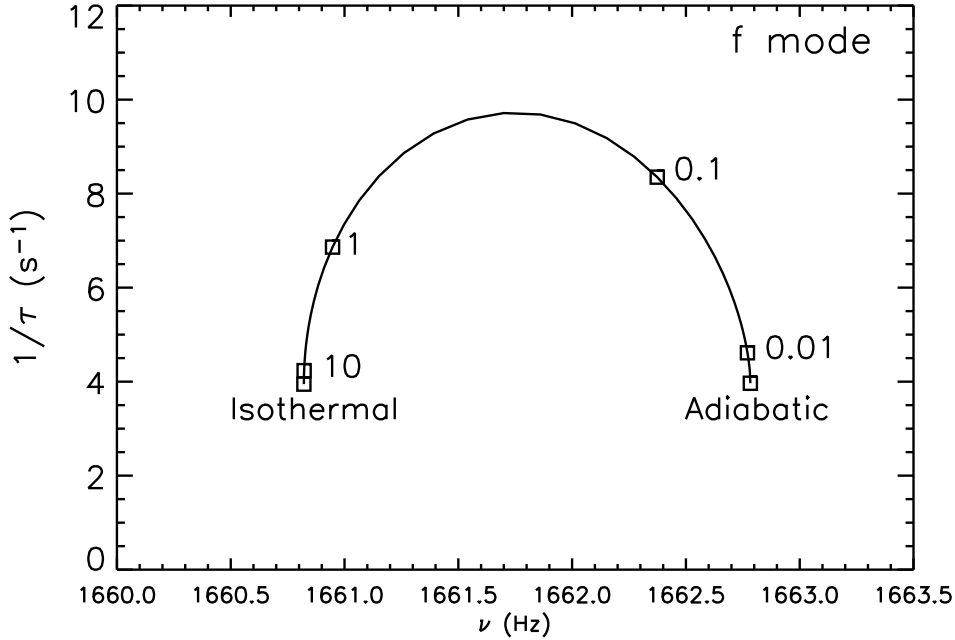} 
\caption{Same as Fig. \ref{figp} but for the fundamental mode.}
\label{figf} 
\end{figure} 
\par 
Let us now focus on the fundamental mode. The $f-$mode is a global
oscillation mode that depends essentially on the average density of
the star, rather than on the velocity at which acoustic waves
propagate. In fact, it is well known that an incompressible fluid
(infinite speed of sound) does not have $p-$modes but the frequency of
the $f-$mode is similar to that of a realistic star with the same
average density.  For these reasons, one expects similar results to
those of the $p-$mode but with smaller effects on the shifts of
frequencies. In Fig. \ref{figf} we show our results for the $f-$mode,
when this is clearly visible. Now, the shift in frequency is only of a
few Hz, and the effect on the damping time is also smaller, although
in the semitransparent regime, it can be as much as a factor 2.5
smaller than that of the adiabatic case (0.25 s). Analogously to what
happened for the $p-$mode, once we enter in the isothermal regime,
thermal damping is less effective and the damping time becomes quite
similar to the adiabatic value.  One must keep in mind that these
results depend on the particular details of the EOS, or more
precisely, on the temperature and specific heat. In cases with higher
temperatures and lower specific heat, the frequency of the fundamental
mode could be quite different from the adiabatic case. It remains to
be analyzed what are the quantitative differences for realistic models
at different stages of neutron star evolution, which is beyond the
scope of this paper.
\par  
\begin{figure}[htb] 
\epsfxsize=14cm \epsfbox{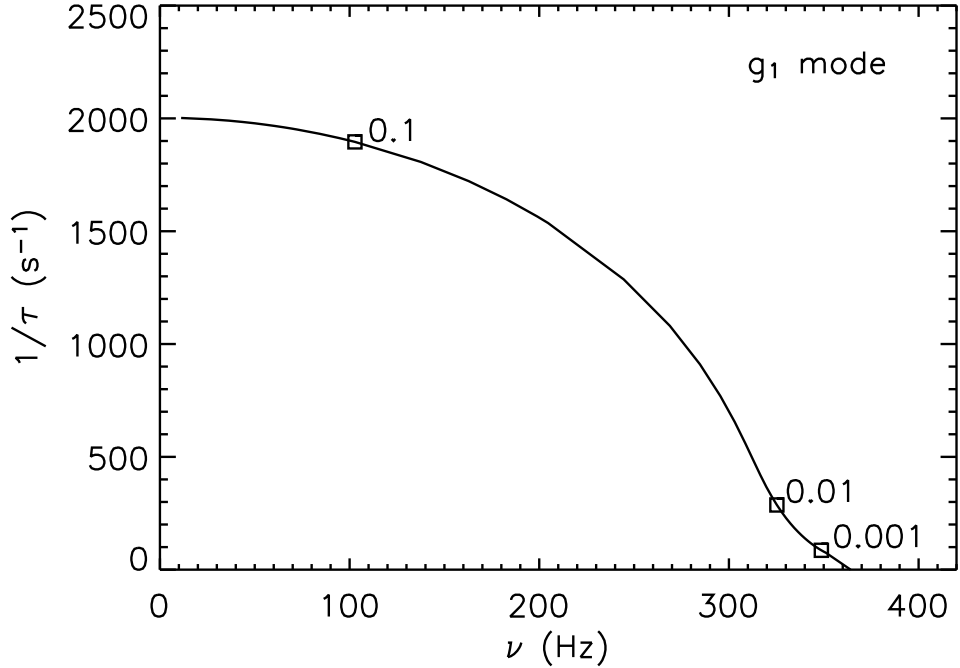} 
\caption{Same as Fig. \ref{figp} but for the first $g-$mode.}
\label{figg} 
\end{figure} 
Finally, let us discuss the $g-$modes. In Fig. \ref{figg} we show the
results for the first gravity mode, which in the adiabatic case has a
frequency of 364 Hz and a damping time of thousands of seconds. The
existence of these modes is related to the presence of thermal (or
chemical) gradients in the star, therefore it is not surprising that
as we increase the thermal conductivity and, consequently, heat
interchange between displaced fluid elements is more effective, the
$g-$mode frequencies decrease and the damping time increases. Contrary
to what happens with the acoustic and fundamental modes, in the limit
of infinite thermal conductivity the $g-$modes degenerate at zero
frequency. It is interesting to note that the damping time does not
decrease indefinitely as the mean free path increases, being bounded
by the minimum dissipative timescale that is, again, close to the
estimate of Eq. \ref{ediss} (of the order of ms), in our scenario.
The last important difference with the other modes is that for values
of $\kappa/n$ as low as $10^{-3}$km, although the change in frequency
is small, the damping time is sensibly shorter (0.01 s). For the $f-$
or $p-$modes it was needed to have higher values of $\kappa/n$ to
obtain significant differences with respect to the adiabatic damping
time.
\section{Final remarks} 
In this paper we have developed the formalism to study non-radial
oscillations of relativistic stars in the frequency domain giving one
step forward from the perfect fluid case by including the effects of
thermal diffusion in a fully relativistic formalism. This allows us to
understand one of the non--adiabatic processes (we did not consider
viscosity) that may be relevant in the study of the oscillation
properties of newly born neutron stars, strange stars, or neutron
stars with deconfined quark matter in the core (hybrid stars).  When a
general equation of state that depends on temperature is used, the
perturbations of the fluid result in perturbations of temperature (or
chemical composition) and, consequently, in heat flux (or chemical
diffusion), that is coupled with the geometry through the Einstein
field equations.  We have analyzed separately the $p-$modes, the
fundamental mode and the $g-$modes, each one being affected in a
different way by thermal diffusion.  As expected from a simplified
model discussed in Appendix \ref{toy}, the $p-$mode frequency is
shifted to lower values in a factor roughly proportional to the ratio
between the {\it isothermal} and {\it adiabatic} speeds of sound
$c_T/c_s$, thus depending very much on the EOS employed and the values
of the local temperature. The frequency of the $f-$mode, however, is
barely affected because it is a global oscillation mode that does not
depend much on the local value of the speed of sound.  Both $f-$ and
$p-$modes are more efficiently damped in the semitransparent regime,
when the mean free path of the particles responsible of the heat
transfer is smaller, but close to the typical length scale of the
system.  The reason is that once the timescale to reach thermal
equilibrium becomes shorter than the oscillation period, the fluid
oscillates keeping the temperature constant, and further increasing
the thermal conductivity does not change this situation. In this
limit, there is no additional thermal damping at first order (as shown
in Appendix \ref{toy}), so that the damping times in the adiabatic and
isothermal limits are very similar.  The response of the $g-$modes to
heat transfer is quite different. Since these modes exist because of
the presence of thermal gradients, it is naturally found that, as heat
transfer becomes more effective, and the temperature perturbations are
smeared out, the frequency is shifted to lower values and the damping
time becomes shorter. In the isothermal limit the $g-$modes have
degenerated to zero frequency.
\par 
In this first approach to the problem, we have only presented results about the
effects of heat transfer. It must be remarked that in newly born neutron stars the effects 
of lepton diffusion are also important, and in Appendix \ref{nm} we discuss the relevant
equations to study that case. The coupling between thermal and chemical diffusion is
one of the complications that arise if one wants to study the realistic case of
proto--neutron stars a few seconds after birth, but there is a second one. In this early
stage the background gradients of temperature and chemical potential are important, 
and in some cases, or for some of the modes, the overall evolution timescale may become
similar to the characteristic oscillation periods or damping times. Furthermore, 
proto--neutron stars are expected to rotate fastly, while in the present work we 
focused on thermal effects neglecting rotation. We defer for future work a more
rigorous study of the realistic scenario, that must consider the non--trivial issue 
of coupling between rotational and thermal effects. This must probably be done
in the time domain instead of the frequency domain, or including second order terms. 
\par 
Our results can also be directly applied
to another interesting case: hot strange stars in which quark matter is in a color
superconductor state. As discussed in the paper, this appears to be the natural state
of deconfined quark matter, and recent calculations show that the thermal conductivities
are many orders of magnitude larger than those of standard neutron star matter. In this
latter scenario, if the strange star is born after the mini--collapse of an old, evolved
neutron star that has been accreting matter, the initial lepton content is small, and
the effects of thermal diffusion are the dominant non--adiabatic correction. 
\acknowledgments
We are indebted to J.A. Miralles and V. Ferrari for many useful comments
and suggestions. We also thank L. Rezzolla and M. Bruni for useful 
discussions. This work has been supported by the Spanish MCyT grant AYA
2001-3490-C02-02, and the {\it Acci\'on Integrada Hispano--Italiana}
HI2003-0284.  J.A.P.~is supported by a {\it Ram\'on y Cajal} contract
from the Spanish MCyT. GM thanks the PPARC for support.
\appendix 
\section{A toy model for non-adiabatic oscillations.} \label{toy} 
Consider a 1-dimensional problem consisting of a box at constant density, 
and constant pressure, and let us study the 
normal acoustic modes of the fluid. The linearized continuity and momentum equations 
in the adiabatic case are 
\beq 
\frac{d\rho_1}{dt} + \rho \nabla v_1 = 0 
\quad, ~ ~ 
\frac{d v_1}{dt} +\frac{1}{\rho} \nabla p_1 = 0 \,.
\eeq 
Integrating in time the continuity equation gives ${\rho_1} + \rho \nabla \xi = 0$, 
with $\xi =\delta r$, 
and substituting ${\rho_1}$ in the momentum equation leads to 
\be 
\frac{d^2 \xi}{dt^2} = -\frac{1}{\rho} \nabla p_1 = 
-\frac{c_s^2}{\rho} \nabla \rho_1 = 
{c_s^2} \nabla^2 \xi 
\label{mom1} 
\ee 
which is a simple wave equation that, when we consider perturbations of the 
form $exp{(i \sigma t - i k x)}$ gives the dispersion equation: 
\be 
- \sigma^2 + k^2 {c_s^2} = 0.
\ee 
Its solution consists of oscillatory modes with frequencies 
$\pm k c_s$. 
\par  
Consider now the non--adiabatic case.  The perturbation of the pressure is 
\be 
p_1 = c_s^2 \rho_1 + \rho \beta s_1 
\ee 
with $\beta= \frac{1}{\rho}(dp/ds)_\rho$, and equation (\ref{mom1}) becomes: 
\be 
\frac{d^2 \xi}{dt^2} = 
{c_s^2} \nabla^2 \xi - {\beta} \nabla s_1 \,.
\label{mom2} 
\ee 
We have to introduce the additional equation of conservation of energy 
\be 
n T \frac{d s}{dt} = - \nabla F 
\ee 
where $T$ is the temperature, $n=\rho/m$ is the particle density, and  
\be 
F =  - \kappa \nabla T \,.
\ee 
\par  
\begin{figure}[htb] 
\epsfxsize=16cm \epsfbox{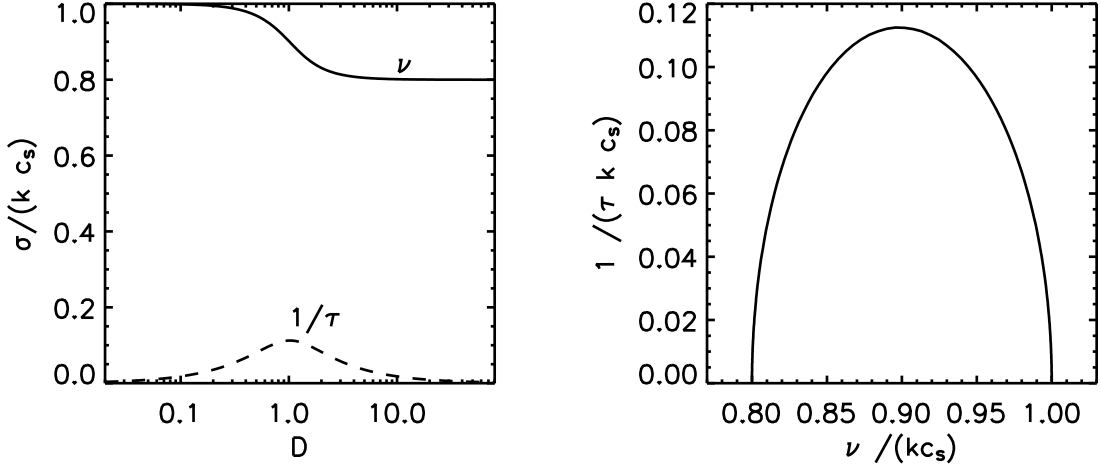} 
\caption{
Left panel: Solution of the dispersion equation
(\ref{disp1}) as a function of $D=\kappa k/n c_v$.The oscillation frequency, 
normalized to $k c_s$ is represented by the solid line, while the inverse 
damping time (imaginary part of the frequency) is denoted by $1/\tau$ (dashed line).
Right panel: representation in the complex plane of the real (frequency) and imaginary
(inverse damping) parts.} 
\label{fappendix} 
\end{figure} 
Considering first order perturbations of the previous equations we have 
\be 
n T \frac{d s_1}{dt} = \kappa \nabla^2 T_1, ~ 
\ee 
that can be used to replace  
$\frac{d s_1}{dt}$ after taking the time derivative of Eq. (\ref{mom2}), to obtain 
\be 
\frac{d^3 \xi}{dt^3} = 
{c_s^2} \nabla^2 \frac{ d\xi}{dt} - \frac{\beta \kappa}{nT} 
\nabla (\nabla^2 T_1)\,. 
\label{mom3} 
\ee 
Next, we only need to find an expression for the last term in Eq. (\ref{mom3}). 
The perturbation of the temperature can be written in terms of $\rho_1$ and $s_1$ 
as follows 
\be 
T_1 = \left(\frac{\partial T}{\partial p}\right)_s c_s^2 \rho_1 + \frac{T}{c_v} s_1~, 
\ee 
where $c_v=T\left(\frac{\partial s}{\partial T}\right)_\rho$ is the specific 
heat at constant volume. Thus ,
\be 
\nabla T_1 = \left(\frac{\partial T}{\partial p}\right)_s c_s^2 \nabla \rho_1 + 
\frac{T}{c_v} \nabla s_1 = - \left(\frac{\partial T}{\partial p}\right)_s \rho c_s^2 
\nabla^2 \xi  - \frac{T}{\beta c_v} \left( \frac{d^2 \xi}{dt^2}- {c_s^2} \nabla^2 
\xi \right) \,.
\ee 
Inserting this in Eq. (\ref{mom3}) one obtains 
\be 
\frac{d^3 \xi}{dt^3} = 
{c_s^2} \nabla^2 \frac{ d\xi}{dt} 
+ \frac{\rho\beta \kappa c_s^2}{nT}\left(\frac{\partial T}{\partial p}\right)_s \nabla^4 \xi 
+ \frac{\kappa}{nc_v} \nabla^2 \left[ 
\frac{d^2 \xi}{dt^2}- {c_s^2} \nabla^2 \xi \right] \,.
\label{mom4} 
\ee 
Considering again perturbations of the form 
$exp{(i \sigma t - i k x)}$ we can derive the dispersion equation: 
\be 
 - \sigma^3  + {c_s^2} k^2 \sigma 
+ i \frac{\kappa}{nc_v} k^2  
\left[ {\sigma^2} -  c_T^2  k^2 \right] 
= 0\,. 
\label{disp1} 
\ee 
The adiabatic limit is recovered when $\kappa /n c_v \ll 1/k$ while 
in the limit $\kappa / n c_v \gg 1/k$, the dispersion 
equation becomes 
\be 
\sigma^2  - k^2 c_T^2 = 0~, 
\ee 
which  simply states that sound waves propagate now at $c_T$ instead of 
$c_s$. It must be remarked that in the isothermal limit, the solution is always  
real (no damping), and  $\sigma$ acquires an imaginary part 
only in the intermediate regime. 
Notice also that dissipation 
is most effective for modes at short wavelengths (large $k$). 
In Fig. \ref{fappendix} we show the solutions of the dispersion equation 
(\ref{disp1}) as a function of the parameter 
$D =\kappa k/nc_v$. The oscillation frequency 
(normalized to $k c_s$) is represented by the solid line, and the inverse 
damping time by the dashed line. The qualitative behaviour is very similar 
to the results discussed in the text. The ratio of the the limiting 
frequencies depends on the ratio $c_T/c_s$. For this example, 
we have taken $c_T/c_s=0.8$. 
\par  
A final remark is in order. The dispersion relation can be written as 
\be 
\gamma^3  + a_2 \gamma^2 + a_1 \gamma_1 + a_0 = 0 
\ee 
\be 
a_2 = D ~~, ~~ 
a_1 = c_s^2 ~~, ~~ 
a_0 = D c_T^2  
\ee 
where $\gamma=i \sigma/ k$. 
The condition that at least one of the roots has a positive real part 
(unstable) is equivalent to one of the following inequalities 
\be 
a_{2} < 0, \;\;\; a_{0} < 0, \;\;\; a_{1} a_{2} < a_{0} \,.
\ee 
Since all coefficients are positive defined, only the third condition 
could apply. It is simply 
\be 
c_s^2 < c_T^2 
\ee 
which is generally not the case, for reasonable realistic EOSs.  
\section{Introducing neutrino diffusion in our equations} 
\label{nm} 
In order to study non-adiabatic perturbations in proto-neutron stars, 
we need to generalize our equations to the case when both, energy flux and 
particle number flux due to neutrino diffusion are present. In this appendix
we derive the equations in a similar way as it was done in the main body
of the paper for the case with only heat transfer. 
\subsection{Thermodynamical variables and relations} 
\label{thermorel}
We choose as independent thermodynamical variables the pressure $p$, the  
entropy per baryon $s$, and the lepton fraction $Y_L=n_L/n$,  
with $n$ being baryon number density  and $n_L$ the lepton number density. 
Given these three thermodynamical quantities, the EOS gives the rest of 
variables. In particular, the EOS provides 
\be
n=n(p,s,Y_L)\,~~~~~\rho=\rho(p,s,Y_L)\,. 
\ee
The reason for taking $Y_L$ as the extensive variable describing the 
chemical composition is the following. The work associated with a 
variation of chemical composition at constant entropy is  
\be \sum_i\mu_idY_i\,.   
\ee  
Here, 
$Y_i=n_i/n$, and $\mu_i,n_i$ are the chemical potentials and number densities 
of the specie $i$, respectively. In our case, the process that changes 
the chemical composition is the capture of electrons by protons and its 
reciprocal (inverse $\beta$ decay)
\be e+p \leftrightarrow n+\nu\,, 
\ee 
so that chemical equilibrium results in $\mu_{\nu}=\mu_e+\mu_p-\mu_n$. 
Since only neutrinos can diffuse throughout the star, we also have 
$dY_e=dY_p=-dY_n$, while $dY_{\nu}$ is independent from the others. 
Consequently,  
\be 
\sum_i\mu_idY_i=(\mu_e+\mu_p-\mu_n)dY_e+\mu_{\nu}dY_{\nu}= 
\mu_{\nu}(dY_e+dY_{\nu})=\mu_{\nu}dY_L\,. 
\ee  
Notice that in this system the variable conjugate to $Y_L$ is $\mu_{\nu}$. 
\par  
According to this result, 
the first principle of thermodynamics (see for example \cite{MTW})  
takes the form 
\be 
d\rho=\frac{\rho+p}{n}dn+nTds+n\mu_{\nu}dY_L\,.\label{first} 
\ee 
Analogously, since we use $(p,s,Y_L)$ as independent variables, we can write, in place  
of (\ref{relalpha}),  
\be 
d\rho=c_s^{-2}dp+nT\alpha_1 ds+n\mu_{\nu}\alpha_2dY_L\label{drho} 
\ee 
where the sound speed $c_s$ is defined by 
\be 
c_s^{2}\equiv\left(\frac{\pa p}{\pa \rho}\right)_{s,Y_L}\,. 
\ee 
and we have defined the following thermodynamical derivatives: 
\be
\alpha_1\equiv\frac{1}{nT}\left(\frac{\pa\rho}{\pa s}\right)_{p,Y_L}~,~~~~~
\alpha_2\equiv\frac{1}{n\mu_{\nu}} 
\left(\frac{\pa\rho}{\pa Y_L}\right)_{p,s}\,. 
\ee 
We also define 
\be 
\alpha_3\equiv\left(\frac{\pa T}{\pa Y_L}\right)_{p,s}= 
\left(\frac{\pa\mu_{\nu}}{\pa s}\right)_{p,Y_L} 
\ee 
where the last equality can be easily proved from Maxwell relations. 
\par 
Let us now consider the heat function $\d Q$ 
\be 
\d Q= Tds +\sum_i\mu_idY_i = Tds+\mu_{\nu}dY_L\,. 
\ee 
Then, we define 
\beq 
\left(\frac{\delta Q}{\pa T}\right)_{p,Y_L}&=&T\left(\frac{\pa s}{\pa T}\right)_{p,Y_L} 
\equiv C_p\nn\\ 
\left(\frac{\delta Q}{\pa\mu_{\nu}}\right)_{p,s}&=&\mu_{\nu} 
\left(\frac{\pa Y_L}{\pa\mu_{\nu}}\right)_{p,s} 
\equiv \Lambda_p\,, 
\eeq 
where $C_p$ is the specific heat at constant pressure and constant number fraction,  
$\Lambda_p$ is a coefficient describing the heat associated to a  
change in composition at constant entropy and pressure.  
With all the above definitions, and using the first law of thermodynamics,  
the expressions of $dT$ and $d\mu_{\nu}$ in terms of $s,p,Y_L$ can be written as  
\beq 
dT&=&\frac{T}{C_p}ds+\frac{T}{\rho+p}(1-\alpha_1)dp+\alpha_3 dY_L\label{dT}\\ 
d\mu_{\nu}&=&\frac{\mu_{\nu}}{\Lambda_p}dY_L 
+\frac{\mu_{\nu}}{\rho+p}(1-\alpha_2)dp+\alpha_3 ds\,.\label{dmu} 
\eeq 
Equations (\ref{drho}), (\ref{dT}), and (\ref{dmu}) allow us to 
express the perturbations of density, temperature and chemical potentials 
in terms of the independent variables. Similar relations apply for
the gradients of the background quantities. For example, from (\ref{drho})
\beq 
\D\rho&=&c_s^{-2}\D p+n\z T\z\alpha_1\D s+n\z\mu_{\nu}\z\alpha_2 
\D Y_L\label{deltarho}\\ 
\rho^{(0)\prime}&=&c_s^{-2}p^{(0)\prime}+nT\z\alpha_1s^{(0)\prime} 
+n\mu_{\nu}\z\alpha_2Y_L^{(0)\prime}\label{rhoprime} 
\eeq 
and the same applies for $\Delta T,\,\Delta\mu_{\nu},\,
T^{(0)\prime},\, \mu_{\nu}^{(0)\prime}$.
\subsection{Dissipative thermodynamics with energy and number transport} 
Next, we need to generalize Eckart dissipative thermodynamics to the case when  
also chemical diffusion, (lepton diffusion in this case) is present. 
Therefore, we need to consider the lepton conservation equation 
\be 
\left(n Y_eu^{\alpha}+ n Y_{\nu}(u^{\alpha}+n^{\alpha})\right)_{;\alpha}=0\,, 
\ee 
where $n^{\alpha}$ is the neutrino four--velocity with respect to 
the frame comoving with matter (see \cite{Thorne}, \cite{IS}), 
with $n^{\mu}u_{\mu}=0$ and $n^{\mu}n_{\mu}=1$. 
Defining the neutrino flux as $f^{\alpha}=n Y_{\nu}n^{\alpha}$, that satisfies 
$u_{\alpha}f^{\alpha}=0\,$, and using the continuity equation (\ref{cont}) 
we obtain the equation of lepton 
conservation $n u^{\alpha} (Y_L)_{,\alpha}=-f^{\beta}_{\,\,;\beta}$. 
This latter equation is particularized for lepton diffusion, which is relevant 
for nascent neutron stars, but it is a general conservation equation for massless 
particles.  Therefore, the set of equations we have is the following: 
\begin{itemize} 
\item Continuity equation: 
\be 
(nu^{\alpha})_{;\alpha}=0\,. 
\ee 
\item Energy conservation equation (stress-energy tensor projected onto $u$): 
\be 
u^\alpha\rho_{,\alpha}+u^{\alpha}_{~;\alpha}(\rho+p)+q^{\alpha}_{~;\alpha} 
+q^{\alpha}u^{\beta}u_{\alpha;\beta}=0\,. 
\ee 
\item First law of thermodynamics: 
\be 
d\rho=\frac{\rho+p}{n}dn+nTds+n\mu_{\nu} dY_L\,.\label{newfirst} 
\ee 
\item Lepton conservation: 
\be 
nu^\alpha Y_{L,\alpha}=-f^{\alpha}_{~;\alpha}\,.\label{Ydot} 
\ee 
\end{itemize} 
Together they give: 
\be 
nTu^\alpha s_{,\alpha}=-q^{\alpha}_{~;\alpha}+\mu_{\nu} f^{\alpha}_{~;\alpha} 
-q^{\alpha}u^{\beta}u_{\alpha;\beta}\, .\label{sdot} 
\ee 
Let us now define the entropy flux as in \cite{Israel},
\be 
S^{\alpha}=snu^{\alpha}+\frac{q^\alpha}{T}-\mu_{\nu}\frac{f^\alpha}{T}\,. 
\ee 
Notice that when degenerate neutrinos dominate the transport of energy
and particles, $q_\alpha=\mu_{\nu} f_\alpha$ and there is no entropy
flux.  By taking the divergence of the entropy flux, we have
\be 
TS^{\alpha}_{~;\alpha}= 
-\frac{q^\alpha}{T}\left(D_{\alpha}T+Tu^\beta u_{\alpha;\beta}\right)
-f^\alpha TD_{\alpha}\eta~, 
\ee 
where we have defined $\eta\equiv\frac{\mu_{\nu}}{T}$.  The second law
of thermodynamics, $S^{\alpha}_{~;\alpha}\ge 0$, must be always
satisfied. The simplest assumption for the form of the energy and
lepton fluxes that satisfies $S^{\alpha}_{~;\alpha}\ge 0$ is
\be
q_\alpha=-\kappa_E(D_\alpha T+Tu^\beta u_{\alpha;\beta})~,~~~~~
f_\alpha=-\kappa_ND_\alpha\eta\label{tqf} 
\ee
with $\kappa_E$, $\kappa_N$ positive defined coefficients, which are,
respectively, the thermal conductivity and the diffusivity governing
the transport of particles.  In the static, radially symmetric case,
our equations coincide with a particular case of those of \cite{Pon99}.
\subsection{Perturbations on a static, radially symmetric, non stratified background} 
Let us consider a static, non--stratified background space--time, in which the
lepton and energy fluxes satisfy 
\beq 
f_{\alpha}\z&=&0\, ~~~~~ q_\alpha\z=0 ~. 
\eeq 
Consistently, we will have  $(Te^{\nu/2})'=0,\,\eta'=0,\,(\mu_{\nu} e^{\nu/2})'=0$.  
Next, we expand the fluxes in spherical harmonics as follows: 
\beq 
\delta q_i&=&\kappa_Ee^{-\nu/2}(Q_{1\,lm}r^{l-1}Y^{lm},Q_{2\,lm}r^lY^{lm}_{,a}) 
e^{\ii\omega t}\nn\\ 
\delta f_i&=&\kappa_N\frac{\eta^2}{(\mu_{\nu} e^{\nu/2})} 
(F_{1\,lm}r^{l-1}Y^{lm},F_{2\,lm}r^lY^{lm}_{,a})e^{\ii\omega t} ~,
\label{hexpansions} 
\eeq 
as well as the Lagrangian perturbation of $Y_L$
\beq
\Delta Y_L&=&-e^{-\nu/2}r^l{\cal Y}_{lm}Y^{lm}e^{\ii\omega t}\,.
\eeq
Analogously to the procedure in subsection 2.1, we define 
\be 
\hat{\cal Y}_{lm}={\cal Y}_{lm}-e^{\nu/2-\lambda} 
\frac{\kappa_E Y_L'}{\ii\omega r}\frac{Q_{1\,lm}}{\rho+p}\,. 
\ee 
By expanding the transport equations (\ref{tqf}), we obtain 
\beq 
\delta q_i&=&-e^{-\nu/2}T\kappa_E\left[\frac{\Delta(Te^{\nu/2})}{T}\right]_{,i} 
-\kappa_ET\omega^2e^{-\nu}v_i\nn\\ 
\delta f_i&=&-\kappa_N\Delta\eta_i\,,\label{transporteqns} 
\eeq  
where $v_j$ has been defined in (\ref{defvj}).
\par 
At this point, we can derive the equations for the perturbations
$\hat E,Q_1,Q_2,F_1,F_2,\hat\Sigma,\hat{\cal Y}$. The differences with
respect to the case with only thermal diffusion are the following:
\begin{itemize} 
\item Expression for $\hat E$. 
\par\noindent
From (\ref{drho}) we have
\par\noindent 
\be 
\hat E_{lm}=c_s^{-2}\hat X_{lm}+nT\alpha_1\hat\Sigma_{lm}+ 
n\mu_{\nu}\alpha_2\hat{\cal Y}_{lm}\,.\label{EqEnumber} 
\ee 
\item Equations for $Q_1$, $F_1$. 
\par\noindent 
Perturbing the conservation equations (\ref{sdot}), (\ref{Ydot}) we have 
\beq 
nT\Delta s+n\mu_{\nu}\Delta Y_L&=&-\frac{e^{\nu/2}}{\ii\omega} 
\left(\delta q^\alpha_{~;\alpha}+\delta q^\alpha u^\beta u_{\alpha;\beta} 
\right)\label{consen}\\ 
n\Delta Y_L&=&-\frac{e^{\nu/2}}{\ii\omega}\delta f^\alpha_{~;\alpha}
\label{consY} 
\eeq 
that, after expanding in harmonics, give 
\beq 
n T \hat\Sigma_{lm}+n\mu_{\nu}\hat{\cal Y}_{lm}&=& 
\kappa_E\frac{e^{\nu/2-\l}}{\ii\omega r} 
\left[Q_{1\,lm}'+\left(\frac{l+1}{r}+\frac{\nu'-\lambda'}{2} 
+\frac{\kappa_E'}{\kappa_E}-\frac{n T s'+n\mu_{\nu}Y_L'}{\rho+p}\right) 
Q_{1\,lm}\right.\nn\\ 
&&\left.-e^{\l}\frac{l(l+1)}{r}Q_{2\,lm}\right]\,,\label{eqSY2}\\ 
n\mu_{\nu}\hat{\cal Y}_{lm}&=&\kappa_N\eta^2\frac{e^{\nu/2-\lambda}}{\ii 
\omega r}\left[F_{1\,lm}'+\left(\frac{l+1}{r}+\frac{\nu'-\lambda'}{2} 
+\frac{\kappa_N'}{\kappa_N}\right)F_{1\,lm}\right.\nn\\ 
&&\left.-e^\lambda\frac{l(l+1)}{r}F_{2\,lm} 
-\frac{n\mu_{\nu}Y_L'}{\rho+p}\frac{Q_1}{\eta^2}\frac{\kappa_E}{\kappa_N} 
\right]\,.\label{expdoty} 
\eeq  
\item Expressions for $\hat\Sigma$, $\hat{\cal Y}$. 
\par\noindent 
Comparing the $i=\theta,\phi$ components of (\ref{hexpansions}) and (\ref{transporteqns})
we obtain
\beq 
e^{-\nu/2}\frac{Q_{2\,lm}}{T}+\omega^2e^{-\nu}\hat V_{lm}
&=& -\left(\frac{\Delta T_{lm}}{T}+\frac{\Delta\nu_{lm}}{2}\right)
\nn\\ 
e^{-\nu/2}\frac{\eta}{\mu_{\nu}}F_{2\,lm}&=& 
-\frac{\Delta\eta_{lm}}{\eta}\,. 
\eeq 
Then, using the thermodynamical relations derived in Section \ref{thermorel},
the Einstein equation (\ref{einsteineqnew}), 
and the fact that the background is static, that leads to
\beq 
\frac{s'}{C_p}-\frac{\alpha_1}{\rho+p}p'+\frac{\alpha_3}{T}Y_L'&=&0\nn\\ 
\frac{Y_L'}{\Lambda_p}-\frac{\alpha_2}{\rho+p}p'+\frac{\alpha_3}{\mu_{\nu}}s' 
&=&0\,,\label{derivatives} 
\eeq 
we find 
\beq 
\hat\Sigma_{lm}+\frac{\alpha_3C_p}{T}\hat{\cal Y}_{lm}&=&C_pA_E\label{EqSigmanumber}\\ 
\hat{\cal Y}_{lm}+\frac{\alpha_3\Lambda_p}{\mu_{\nu}}\hat\Sigma_{lm}&=& 
\Lambda_pA_N\label{EqYnumber} 
\eeq 
where 
\be
A_E=\frac{\alpha_1}{\rho+p}\hat X_{lm}+\frac{Q_{2\,lm}}{T}~,~~~~~
A_N=\frac{\alpha_2}{\rho+p}\hat X_{lm}+\frac{F_{2\,lm}+Q_{2\,lm}}{T}\,. 
\ee 
The solution of this system is 
\beq 
\hat\Sigma_{lm}&=&\frac{C_p}{1-\frac{\alpha_3^2C_p\Lambda_p}{T\mu_{\nu}}} 
\left[A_E-\frac{\alpha_3\Lambda_p}{T}A_N\right]\nn\\ 
\hat{\cal Y}_{lm}&=&\frac{\Lambda_p}{1- 
\frac{\alpha_3^2C_p\Lambda_p}{T\mu_{\nu}}} 
\left[A_N-\frac{\alpha_3C_p}{\mu_{\nu}}A_E\right]\,. 
\eeq 
\item Equations for $Q_2,F_2$. 
\par\noindent 
The equations for $Q_2$, $F_2$ follow from the fact that  
$e^{\nu/2}\delta q_j/\kappa_E+T\omega^2e^{-\nu/2}v_j$ and  
$\delta f_j/\kappa_N$ are gradients, which can be used to relate the
radial and angular components of (\ref{hexpansions}), 
\beq
F_{2\,lm}'&=&-\frac{l}{r}F_{2\,lm}+\frac{1}{r}F_{1\,lm}\,.\label{EqF2number} 
\\
Q_{2\,lm}'&=&-\frac{l}{r}Q_{2\,lm}+\frac{1}{r}Q_{1\,lm} 
+\frac{T}{(\rho+p)^2}(\hat X_{lm}\rho'-\hat E_{lm}p')\,.
\label{EqQ2n} 
\eeq 
The last term in (\ref{EqQ2n}) can be expanded in terms of the rest of variables.
Using (\ref{rhoprime}) and (\ref{EqEnumber}), and after some algebra, 
one finds 
\beq 
Q_{2\,lm}'&=&-\frac{l}{r}Q_{2\,lm}+\frac{1}{r}Q_{1\,lm}\nn\\ 
&&+\frac{\nu'}{2(\rho+p)}\frac{1}{1-\frac{\alpha_3^2C_p\Lambda_p}{T 
\mu_{\nu}}}\left\{nT\alpha_1C_p\left[\left(1-\frac{\alpha_3\Lambda_p}{T}\right) 
Q_{2\,lm}-\frac{\gamma_2\Lambda_p}{T}F_{2\,lm}\right] 
\right.\nn\\ 
&&\left.+n\mu_{\nu}\alpha_2\Lambda_p 
\left[\left(1-\frac{\alpha_3C_p}{\mu_{\nu}}\right) 
Q_{2\,lm}+F_{2\,lm}\right]\right\}\,.\label{EqQ2number} 
\eeq 
\end{itemize} 
Equations (\ref{EqEnumber}), (\ref{eqSY2}), (\ref{expdoty}), (\ref{EqSigmanumber}),  
(\ref{EqYnumber}), (\ref{EqF2number}), (\ref{EqQ2number}), together with equations  
(\ref{exH0}) to (\ref{eqX}), form a closed system  
that allows to compute the non-radial perturbations of a star with  
energy and lepton transport. 
 
\end{document}